\documentstyle[epsfig]{cupconf}
\title[MOND and its implications] {Modified Newtonian Dynamics and its
Implications}

\author[R.H. Sanders] {R.\ns H.\ns S\ls A\ls N\ls D\ls E\ls R\ls S}

\affiliation{Kapteyn Astronomical Institute, Groningen, The Netherlands}

\begin{document}

\maketitle

\begin{abstract}
Milgrom has proposed that the appearance of discrepancies
between the Newtonian dynamical mass and the directly observable
mass in astronomical systems could be due to a breakdown of Newtonian
dynamics in the limit of low accelerations rather than the presence of
unseen matter.  Milgrom's hypothesis, modified Newtonian dynamics or
MOND , has been remarkably successful in explaining systematic
properties of spiral and elliptical galaxies and predicting in
detail the observed rotation curves of spiral galaxies with only
one additional parameter-- a critical acceleration which is
on the order of the cosmologically interesting value of $cH_o$.
Here I review the empirical successes of this idea and discuss
its possible extention to cosmology and structure formation.
\end{abstract} 

\section{Introduction}

Modified Newtonian dynamics (MOND) is an {\it ad hoc} modification of 
Newton's law of gravity or inertia
proposed by Milgrom (1983) as an alternative to cosmic dark matter.  
The motivation for this and other such proposals is obvious:  So long 
as the only evidence for dark
matter is its global gravitational effect, then its presumed existance
is not independent of the assumed form of the law of gravity or inertia on 
astronomical scales.  In other words, either the universe contains large
quantities of unseen matter, or gravity (or the response of particles to
gravity) is not generally the same as it appears to be in the solar system.

The phenomenological foundations for MOND really come down to two
observational facts about spiral galaxies: 1.)  The rotation curves
of spiral galaxies are asymptotically flat, and 2.)  There is a well-defined
relationship between the rotation velocity in spiral galaxies and the
luminosity-- the Tully-Fisher (TF) law (\cite{tf77}).  This latter 
implies a mass-velocity   
relationship of the form $M\propto V^\alpha$ where $\alpha$ is in the
neighborhood of 4.  

If one wants to modify gravity in some way to explain 
flat rotation curves or the existance of a mass-rotation velocity relation 
for spiral galaxies, an
obvious first choice would be to propose that gravitational attraction becomes
more like 1/r beyond some length scale which is comparable to the scale of
galaxies.  So the modified law of attraction about a point mass M would read
$$F = {GM\over{r^2}}{f(r/r_o)}\eqno(1)$$ where $r_o$ is a new constant of 
length with dimensions of a few kpc, and $f(x)$ is a function with the
asymptotic behavior: $f(x)=1$ where $x<<1$ and $f(x)=x$ where $x>>1$. 
Equating the centripetal to the gravitational
acceleration in the limit $r>>r_o$ 
would lead to a  mass-- asymptotic rotation velocity relation of the form
$v^2 = GM/r_o$. This is true of any modification attached to a length scale. 
Milgrom realized that this was incompatible with the observed TF law unless,
of course, the mass-to-light ratio (M/L) of the stellar population 
varies systematically with galaxy mass in a very 
dramatic fashion.  Such a drastic variation in M/L ($\propto M^{-2}$),
is absolutely inconsistent with everything we think we know about 
stellar populations.  Moreover, any modification attached to a length
scale would imply that larger galaxies should exhibit a larger discrepancy.
Anyone who has considered galaxy rotation curves knows that this is 
totally inconsistent with the observations.  There are very small, usually
low surface brightness (LSB) galaxies with large discrepancies, and very 
large high surface brightness (HSB) spiral
galaxies with very small discrepancies.  

This is shown in the
first figure.
\begin{figure}
\begin{center}
 \epsfig{file=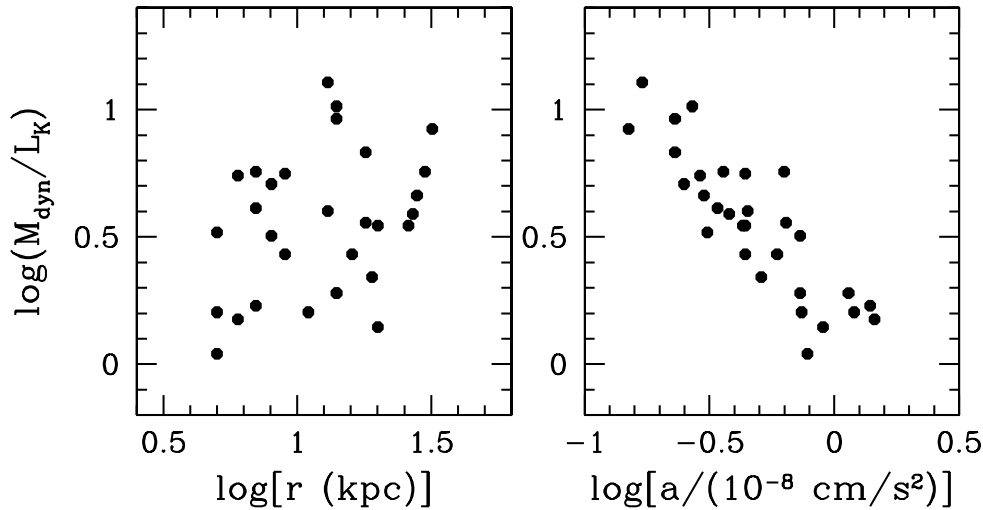,width=14cm}\\
 \caption{The global mass-to-K' band luminosity of Ursa Major spirals at the
last measured point of the rotation curve plotted first against the
radial extent of the rotation curve (left) and then against the centripetal
acceleration at that point (right).}
\end{center}
\end{figure}
At the left is a log-log plot of the dynamical $M/L_{K'}$
vs.\ the radius at the last measured
point of the rotation curve for a uniform sample of spiral 
galaxies in the Ursa Major cluster (Tully et al. 1996, 
Verheijen and Sancisi, 2001).  The dynamical
M/L is calculated simply using the Newtonian formula for the mass $v^2r/G$ 
(assuming a spherical mass distribution) where r is the 
radial extent of the rotation curve.
Population synthesis  studies suggest that $M/L_{K'}$ should be about one,
so anything much above one indicates a discrepancy-- a dark matter problem.  
It is evident that there is not much of a correlation of M/L with size.  
On the other
hand, the Newtonian M/L plotted against centripetal acceleration ($v^2/r$)
at the last measured point (right figure) looks rather different.  
There does appear
to be a correlation in the sense that $M/L\propto 1/a$ for $a<10^{-8}$
cm/s$^2$.  Any modification of gravity 
attached to a length scale cannot explain such observations.

\section{Basics of MOND}

Milgrom's insightful deduction was that the only viable sort of modification
is one in which a deviation from Newton's law appears at low acceleration.
(it should be recalled that data such as that shown in Fig.\ 1 did not exist
at the time of Milgrom's initial papers).  Viewed as a modification
of gravity, his suggestion was that the actual gravitational
acceleration ${\bf g}$ is related to the Newtonian gravitational acceleration
${\bf g_n}$ as
$$ {\bf g}\mu(|g|/a_o) = {\bf g_n}\eqno(2)$$ where 
$a_o$ is a new physical parameter with
units of acceleration and $\mu(x)$ is a function which is unspecified
but must have the asymptotic form $\mu(x) = x$ when $x<<1$ and 
$\mu(x) = 1$ where $x>>1$. 

The immediate consequence of this is that, in the limit of low 
accelerations, $g = \sqrt{g_na_o}$.  For a point mass M, if we set g equal
to the centripetal acceleration $v^2/r$, this gives
$$v^4 = GMa_o\eqno(3)$$ in the low acceleration regime.  
So all rotation curves are
asymptotically flat and there is a mass-luminosity relation of the form
$M\propto v^4$.  These are aspects that are built into MOND so they
cannot rightly be called predictions.  However,
in the context of MOND, the  aspect of an asymptotically flat rotation curve 
is absolute.  MOND leaves rather little room for maneuver;  the idea is 
in principle
falsifiable, or at least it is far more fragile than the 
dark matter hypothesis.  Unambiguous examples of rotation curves 
(of isolated galaxies) which 
decline in a Keplerian fashion at a large distance
from the visible object would falsify the idea. 
In effect, a rotational velocity which is constant with radius is
Kepler's law in the limit of low accelerations.

In addition, the mass-rotation velocity relation and implied Tully-Fisher
relation is absolute.  The
TF relation should be the same for different classes of galaxies and the
logarithmic slope (at least of the MASS-velocity relation) must be 4--
not 3.8 or 4.2-- but 4.0.  Moreover, it must be the case that 
the relation is essentially one between the total
baryonic mass of a galaxy and the asymptotic flat rotational velocity--
not the peak rotation velocity but the velocity at large distance.
This is the most immediate and most obvious prediction (see McGaugh \&
de Blok 1998b and McGaugh et al. 2000 for a discussion of these points).  

Converting the M-V relation 
to the observed luminosity-velocity relation we find
$$ log(L) = 4log(V) - log(Ga_o<M/L>).\eqno(4)$$
The near-infrared TF relation for Verheijen's UMa sample is shown
in Figure 2 (\cite{sv98}) where the velocity is that of
the flat part of the rotation curve.  The scatter about the 
least-square fit line of slope $3.9 \pm 0.2$ is consistent with observational
uncertainties (i.e., no intrinsic scatter).
\begin{figure}
\begin{center}
\epsfig{file=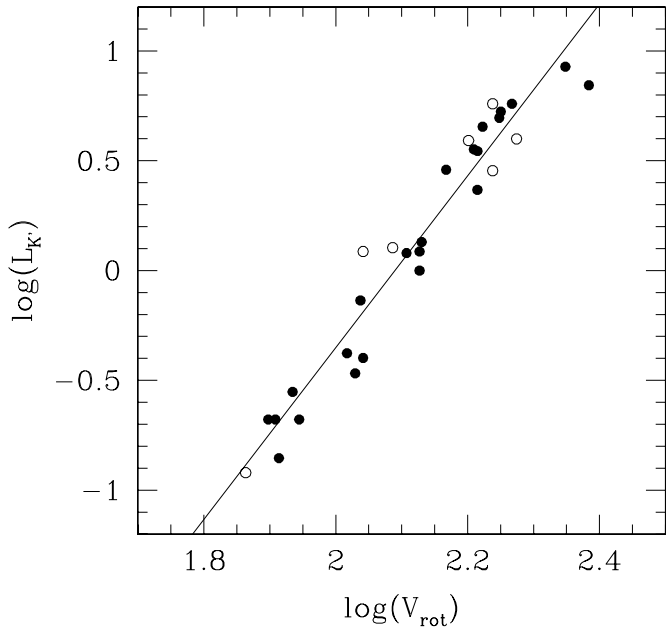,width=10cm}\\
\caption{The near-infrared Tully-Fisher relation of Ursa Major spirals 
(\cite{sv98}).  The rotation velocity is the asymptotically constant value.
The line is a least-square fit to the data and has a slope of $3.9\pm 0.2$}
\end{center}
\end{figure}
Given the mean M/L in a particular band ($\approx 1$ in the 
K' band), this observed TF relation (eq.\ 4) tells us that
$a_o$ must be on the order of $10^{-8}$ cm/s$^2$.  
It was immediately noticed by Milgrom
that $a_o\approx cH_o$ to within a factor of 5 or 6.
This cosmic coincidence is quite interesting and suggests that MOND, if it
is right, may reflect the effect of cosmology on local particle dynamics.

\section{Implications}

There are several other immediate consequences of modified dynamics--
all of which were explored by Milgrom in his original papers-- which
do fall in the category of predictions.

1.  There exist a critical value of the surface density
$$\Sigma_c \approx a_o/G.\eqno(5)$$  If a system, such as a spiral galaxy has
a surface density of matter greater than $\Sigma_c$, that means
that the internal accelerations are greater than $a_o$, so the system is
in the Newtonian regime.  In systems with $\Sigma \geq \Sigma_c$ (HSB
galaxies) there should be a small discrepancy between
the visible and classical Newtonian dynamical mass within the optical disk. 
In the parlance of rotation curve observers, a HSB galaxy should be
well-represented by the ``maximum disk'' solution (Sancisi, this volume).
But in LSB galaxies ($\Sigma<<\Sigma_c$) there is a low internal 
acceleration, so the discrepancy between the visible
and dynamical mass would be large.  These objects should be far from
maximum disk.  In effect, Milgrom predicted, before the actual
discovery of LSB galaxies, that there would be a serious
discrepancy between the observable and dynamical mass within the
luminous disk of such systems-- 
should they exist.  They do exist, and
this prediction has been verified-- as is evident from the work of
McGaugh \& de Blok (1998a,b).

2.  It is well-known since the work of Ostriker \& Peebles (1973), that 
rotationally supported Newtonian systems tend to be unstable to
global non-axisymmetric modes which lead to bar formation and rapid
heating of the system.  In the context of MOND, these systems would be
those with $\Sigma > \Sigma_c$, so this would suggest that $\Sigma_c$
should appear as an upper limit on the surface density of rotationally 
supported systems.  This critical surface density is
0.2 g/cm$^2$ or 860 M$_\odot$/pc$^2$.  A more appropriate value of the mean
surface density within an effective radius would be $\Sigma_c/2\pi$
or 140 M$_\odot/pc^2$, and, taking
$M/L_b \approx 2$, this would correspond to a surface brightness of
about 22 mag/arc sec$^2$.  There is 
such an observed upper limit on the mean surface brightness of spiral 
galaxies and this is known as
Freeman's law (Freeman 1970, Allen \& Shu 1979).  The point is that 
the existance of such a preferred surface density
becomes understandable in the context of MOND.

3.  Spiral galaxies with a mean surface density near this limit -- 
HSB galaxies-- would be, within
the optical disk, in the Newtonian regime.  So one would expect that
the rotation curve would decline in a near Keplerian fashion to the 
asymptotic constant value.  In LSB galaxies, with mean surface density
below $\Sigma_c$, the prediction is that rotation curves would rise to
the final asymptotic flat value.  So there should be a general difference
in rotation curve shapes between LSB and HSB galaxies.
In Fig.\ 3 I show the rotation curves of two galaxies, a LSB and HSB, 
where we
see exactly this trend.  This general effect in observed rotation curves
was first noted by Casertano \& van Gorkom (1991).
\begin{figure}
\begin{center}
\epsfig{file=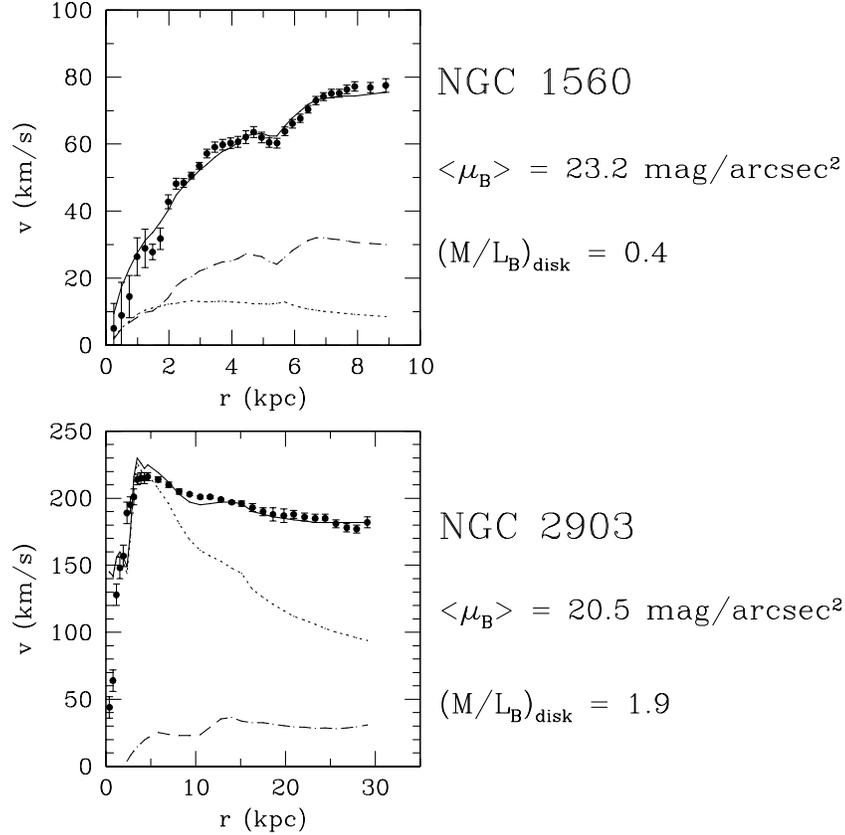,width=12cm}
\caption{The points show the observed 21 cm line rotation curves of
a low surface brightness galaxy, NGC 1560 (Broeils 1992) and a high
surface brightness galaxy, NGC 2903 (Begeman 1987).  The dotted and dashed
lines are the Newtonian rotation curves of the visible and gaseous
components of the disk and the solid line is the MOND rotation curve 
with $a_o=1.2\times 10^{-8}$ cm/s$^2$-- the value derived from the
rotation curves of 10 nearby galaxies (Begeman et al. 1991).  Here
the only free parameter is the mass-to-light ratio of the visible 
component.}
\end{center}
\end{figure}

4.  With Newtonian dynamics, pressure-supported systems which 
are nearly isothermal have infinite extent.  But in the context of MOND
it is straightforward to demonstrate that such isothermal systems
are finite with the density at large radii falling roughly like $1/r^4$
(Milgrom 1984).  
The equation of hydrostatic equilibrium for an isotropic, isothermal
system reads $${\sigma_r}^2 {{d\rho}\over{dr}} = -\rho g\eqno(5)$$
where, in the limit of low accelerations $g=\sqrt{GMa_o}/r$.
Here $\sigma_r$ is the radial velocity dispersion and $\rho$ is
the mass density.
It then follows immediately that, in this MOND limit,
$$\sigma_r^4 = {GMa_o\Bigl({{d\,ln(\rho)}\over{d\,ln(r)}}\Bigl)^{-2}}.
\eqno(6)$$
Thus there exists a mass-velocity dispersion
relation of the form $$(M/10^{11}M_\odot) \approx (\sigma_r/100\,\,
kms^{-1})^4$$ which is similar to the observed
Faber-Jackson relation (luminosity-velocity dispersion relation) 
for elliptical galaxies (\cite{fj}).   
This means that a MOND near- isothermal sphere 
with a velocity dispersion of 100 km/s to
300 km/s will always have a galactic mass.  This is not 
true of Newtonian pressure-supported objects.  Because of the appearance
of an additional dimensional constant, $a_o$, in the structure equation 
(eq.\ 5), MOND systems are much more constrained than their Newtonian
counterparts.
 
But with respect to actual pressure supported systems, an even stronger
statement can be made.  Any isolated system which is nearly isothermal will 
be a MOND object.  That is because a Newtonian isothermal system  
(with large internal accelerations) 
is an object of infinite size and will always extend to the
region of low accelerations ($<a_o$).  At that point (${r_e}^2\approx 
GM/a_o$), MOND intervenes and the system will be truncated.
This means that the internal acceleration of any isolated 
isothermal system (${\sigma_r}^2/r_e$) is expected to be on the
order of or less than $a_o$ and that the mean surface density 
within $r_e$ will typically be $\Sigma_c$ or less 
(there are low-density solutions for MOND isothermal spheres,
$\rho<<{a_o}^2/G\sigma^2$, with internal accelerations less than $a_o$).  
It has been known for some time that elliptical galaxies
do have a characteristic surface brightness (Fish 1964).
But the above arguments imply that the same should be true of any
pressure supported, near-isothermal system, from globular clusters
to clusters of galaxies.  Moreover, the same $M-\sigma$ relation 
(eq.\ 6) should apply to all such systems, albeit with considerable
scatter due to deviations from a strictly isotropic, isothermal
velocity field (Sanders 2000).  Such deviations will also result in
a dispersion of mean internal accelerations about the fiducial value of
$a_o$.

\section{Rotation curve analysis}

Perhaps the most remarkable phenomenological success of MOND is in predicting 
the form of rotation curves from the observed distribution of
detectable matter-- stars and gas (Begeman et al. 1991, McGaugh \&
de Blok 1998, Sanders \& Verheijen 1998).  The procedure followed can be
outlined as follows:

1.  One assumes that light traces mass, i.e., M/L = constant.  There
are color gradients in spiral galaxies so this cannot be generally
true-- or at least one must decide which color band is the best tracer
of the mass distribution.  The general opinion is that
the near-infrared emission of spiral galaxies is the optimal tracer
of the underlying stellar mass distribution, since the old population of
low mass stars
contribute to this emission and the near-infrared is less affected by dust
obscuration.  So where available, near infrared surface photometry is to
be preferred.

2.  In determining the distribution of detectable matter one must 
include the observed neutral hydrogen with an appropriate correction for the
contribution of primordial helium.  The gas can make a dominant
contribution to the total mass surface density in some (generally low
luminosity) galaxies.  

3.  Given the observed distribution of mass, $g_n$, the Newtonian gravitational
force, is calculated via the  classical Poisson
equation.  Here it is usually assumed that the stellar and gaseous disks
are razor thin.  It may also be necessary to add a spheroidal bulge 
if the light distribution indicates the presence of such a component.

4.  Given the radial distribution of the Newtonian force, the true
gravitational force, $g$, is calculated from
the MOND formula with $a_o$ fixed.  Then the mass of the stellar disk is
adjusted until the best fit to the observed rotation curve is achieved.  This
gives M/L of the disk as the single free parameter of the fit (unless a bulge
is present).

In comparing to the observed rotation curve one assumes that the motion of 
the gas is co-planer rotation about the center of the given galaxy.  
This is certainly not always the case because there are well-known 
distortions to the velocity field in spiral galaxies caused by bars and 
warping of the gas layer.  In a fully 2-dimensional velocity field these 
distortions can often be modeled, but the optimal rotation curves are those 
in which there is no evidence for the presence of significant deviations 
from co-planer circular motion.  In general it should be remembered that
not all observed rotation curves are perfect tracers of the radial
distribution of force.  A perfect theory will not fit all rotation curves
because of these possible problems (the same is true of a specified
dark matter halo).
The point is that with MOND, usually, there is one free parameter per 
galaxy and that is the mass or M/L of the stellar disk.

I am only going to show two examples of MOND fits to rotation curves,
and these are the two galaxies already shown in Fig.\ 3.  The dotted and
dashed curves are the Newtonian rotation curves of the stellar and gaseous
disks respectively, and the solid curve is the MOND rotation curve with 
$a_o = 1.2\times 10^{-8}$ cm/s$^2$.  We see that, not only does MOND 
predict the general trend for LSB and HSB galaxies, but it also predicts
the observed rotation curves {\it in detail} from the observed distribution
of matter.  This procedure has been carried out for about 100 rotation 
curves and in only about 10 cases is the predicted rotation curve 
significantly different from the observed curve.  For these 
objects there is usually an obvious problem with the observed curve or
its use as a tracer of the radial force distribution.
  
I have noted that the only free parameter in these fits is the 
mass-to-light ratio of the visible disk, so one may well ask
if the inferred values are reasonable.  Here it is useful to consider
again the Verheijen UMa sample because
all galaxies are at the same distance and there is K'-band (near infrared)
surface photometry of the entire sample.  The sample also contains both
HSB and LSB galaxies.  Fig.\ 5 shows the M/L in the B-band
required by the MOND fits 
plotted against B-V color (top) and the same for the K'-band (bottom).
We see that in the K'-band M/L $\approx 1$ with a 30\% scatter.
In other words, if one were to assume a K'-band M/L of one at the outset, 
most rotation 
curves would be quite precisely predicted from the observed light and gas
distribution with no free parameters.
In the B-band, on the other hand, the MOND M/L does appear to be a function
of color in the sense that redder objects have larger M/L values.
This is exactly what is expected from population synthesis models
as is shown by the solid lines in both panels (Bell \& de Jong 2000).
This is quite interesting because there is nothing built into MOND
which would require that redder galaxies should have a higher $M/L_b$;  this
simply follows from the rotation curve fits.

\begin{figure}
\begin{center}
\epsfig{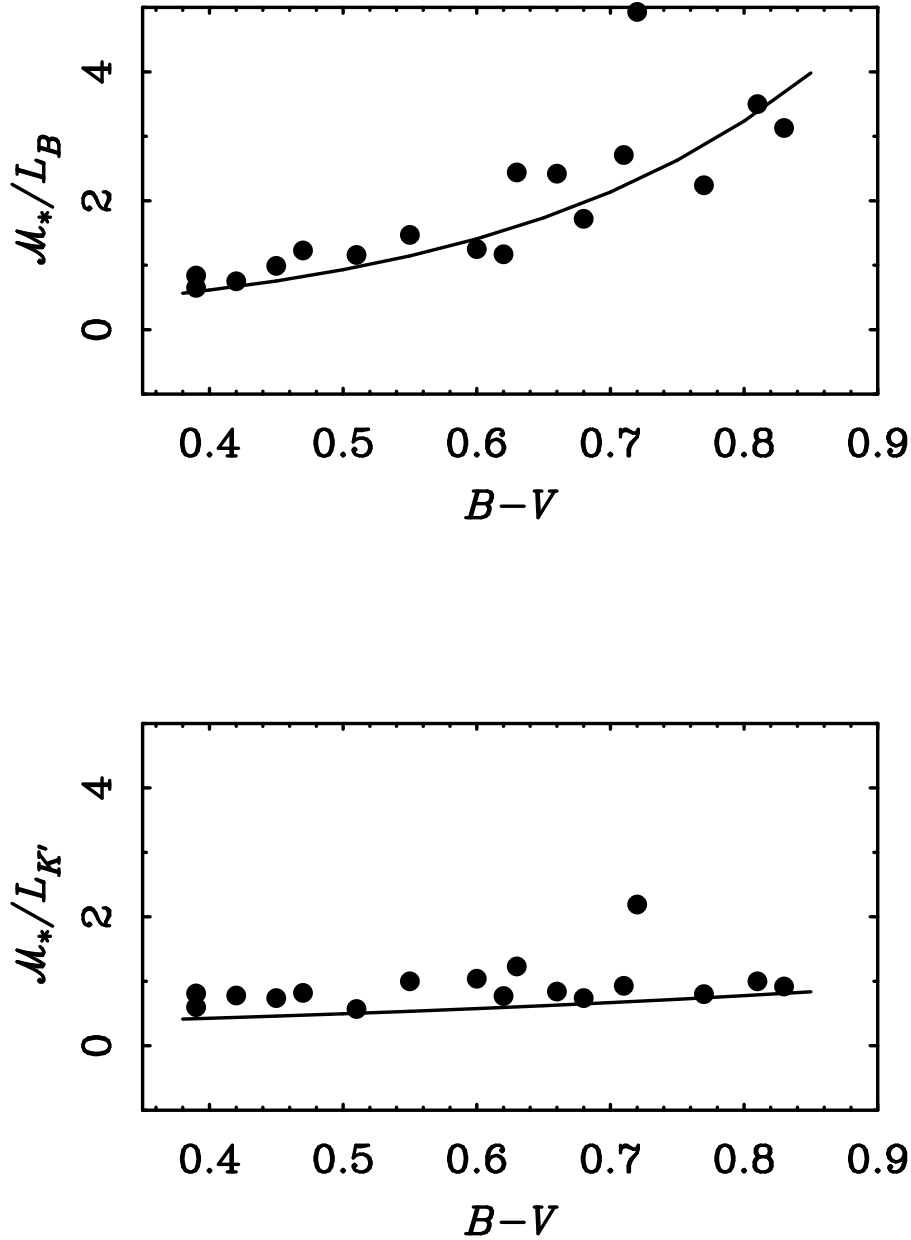}
\caption{Inferred mass-to-light ratios for the UMa spirals (Sanders
\& Verheijen) in the B-band (top) and the K'-band (bottom)
plotted against B-V colors (McGaugh, private communication).  
The solid lines show predictions from
populations synthesis models by Bell and de Jong (2001).}
\end{center}
\end{figure}

We sometimes hear that it is not so surprising that MOND fits rotation
curves because that is what it was designed to do.  This is certainly not
correct.  MOND was designed to produce asymptotically flat rotation curves
with a given mass-velocity relation (or TF law).  It was not designed to
fit the details of all rotation curves with a single adjustable parameter
(even of galaxies which are gas-dominated with no adjustable parameter), 
and it was certainly not designed to provide a reasonable 
dependence of fitted M/L on color.  Indeed, there are a couple of 
well-observed
spiral galaxies which are problematic for MOND, and which could, in 
principle, falsify the idea.  One of these is NGC 2841--
a large spiral galaxy with a Hubble distance of about 9 Mpc (Begeman et al.
1991).  
\begin{figure}
\begin{center}
\epsfig{file=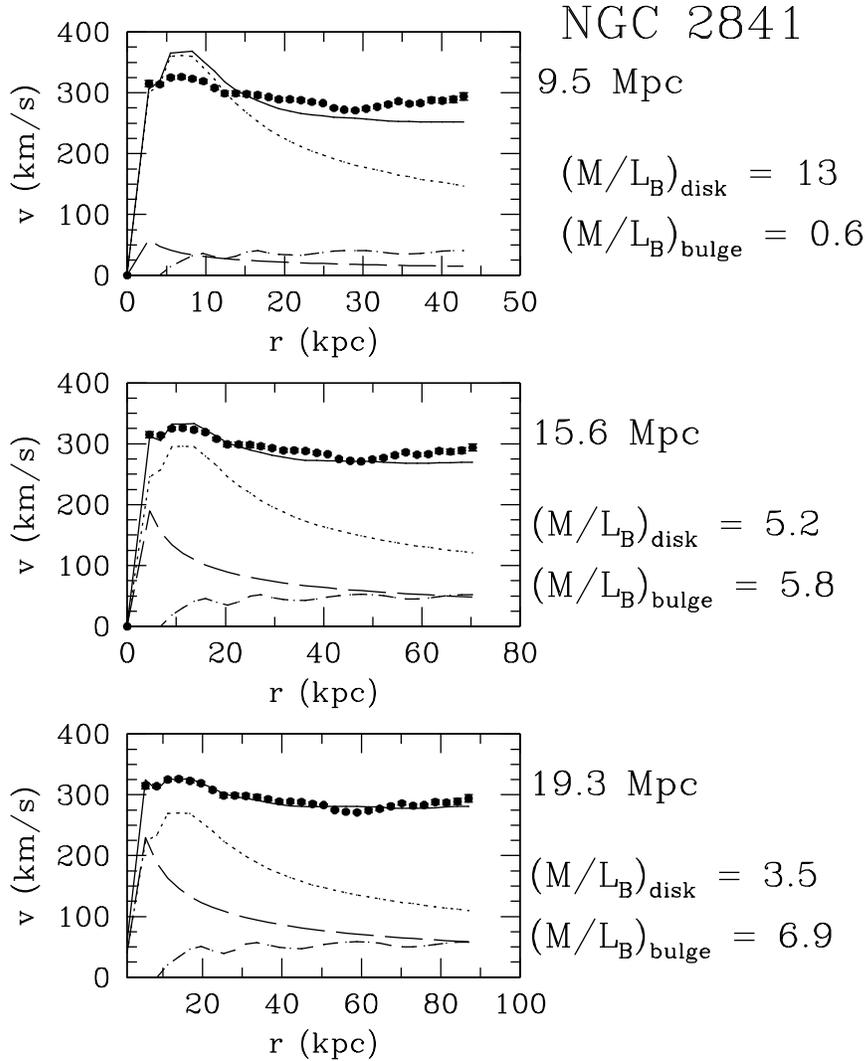,width=15cm}
\caption{MOND fits to NGC 2841 at various distances.  The Hubble law distance
is 9.3 Mpc (h=0.75), but MOND prefers a distance of 19.3 Mpc.  The 
Cepheid distance is $14.1\pm 1.5$.  The MOND rotation curve at the Cepheid 
distance +1$\sigma$ (15.6 Mpc) is acceptable, particularly considering the
complication of the large warp in the outer regions}
\end{center}
\end{figure}
In fact, the rotation curve of the galaxy cannot be
fit using MOND if the distance is only 9 Mpc.  MOND prefers a distance
of 19 Mpc as we see in Fig.\ 5 (the scaling of the centripetal acceleration
depends upon the distance).  If the distance to
this galaxy is really less than about 14 Mpc it is quite problematic for MOND.
Now it turns out that a Cepheid distance to this galaxy
has just been determined (\cite{maceal01}), and this is 14.1 $\pm$ 1.5 Mpc.  
Given that the distance could easily be as large
as 15.6 Mpc, this galaxy now would seem to present no problem for MOND.

The success of MOND in accounting for galaxy rotation curves with
only one free parameter, the M/L of the visible disk which usually
assumes quite reasonable values, is remarkable.  {\it Whether
MOND is correct or not, the success of this simple algorithm 
implies that galaxy rotation curves
are entirely determined by the distribution of visible matter.  If
you believe in dark matter, then you somehow must explain this
phenomenology.  How can the distribution of 
dark matter be so intimately connected with
the distribution of visible matter?}

\section{Pressure-supported systems}

\begin{figure}
\begin{center}
\epsfig{file=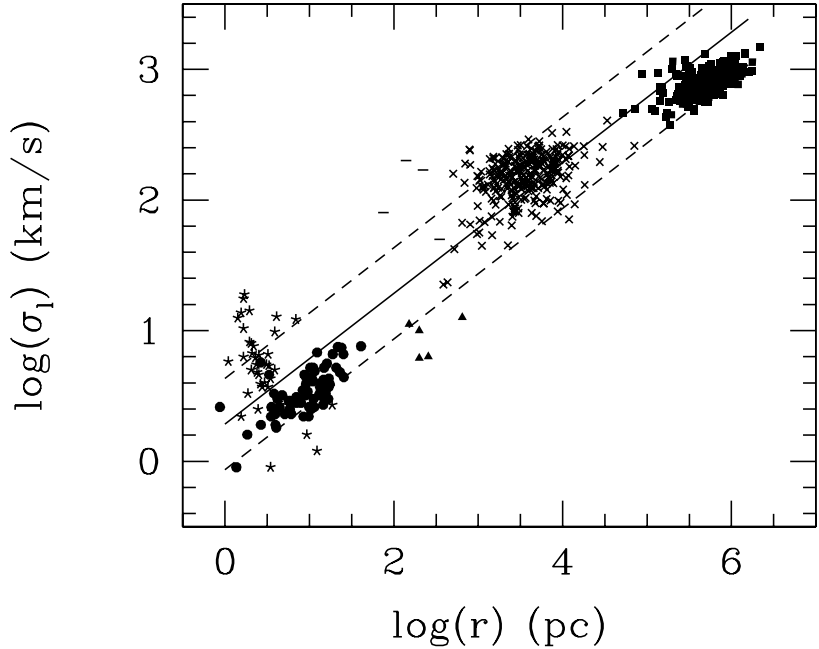,width=12cm}
\caption{The line-of-sight velocity dispersion vs. characteristic
radius for pressure-supported astronomical systems.  The star-shaped
points are globular clusters (Trager et al. 1993), the points are
massive molecular clouds in the Galaxy (Solomon et al. 1987), the
crosses are massive elliptical galaxies (Jorgensen et al. 1995a,b),
and the squares are X-ray emitting clusters of galaxies (White, Jones
\& Forman 1997).  The solid line is shows the relation 
$\sigma_l^2/r = a_o$ and the dashed lines a factor of 5 variation about
this relation.}
\end{center}
\end{figure}

Fig.\ 6 is a log-log plot
of the velocity dispersion versus size for pressure-supported, 
nearly isothermal astronomical systems.  At the bottom of the plot are 
globular clusters (star-shaped) and giant molecular clouds (points) in the
Galaxy.  The group of points 
in the middle are ellipticals (crosses) and at the top are X-ray emitting 
clusters of galaxies (squares).  The triangles are the dwarf spheroidal 
systems surrounding the Milky Way and the dashes are compact dwarf 
ellipticals.  The plotted parameters have not been 
massaged at all but are taken directly from the relevant observational papers. 
The  measure of size is not homogeneous-- for ellipticals and globular clusters
it is the well-known effective radius, for the X-ray clusters it is an
X-ray intensity isophotal radius, and for the molecular clouds it is a
isophotal radius of CO emission.  The velocity dispersion refers to the
central velocity dispersion for ellipticals and globulars; for the clusters
it is the thermal velocity dispersion of the hot gas;  for the molecular 
clouds it is just the typical line width of the CO emission.  A 
velocity-dispersion-- size correlation
has been previously claimed for individual classes of objects-- 
most notably, the molecular clouds and the clusters of galaxies.

The parallel lines are not fits but represent
fixed internal accelerations.  The solid line corresponds to 
${\sigma_l}^2/r = 10^{-8}$ cm/s$^2$ and the 
parallel dashed lines to accelerations 5 times larger or smaller
than this particular value.  It is clear from this diagram
that the internal accelerations in these systems all lie within a factor
of a few of $a_o$.  This also implies that the surface densities in
these systems are near the MOND surface density $\Sigma_c$.  

So these astronomical objects appear to have a characteristic internal
acceleration or a characteristic surface density as MOND predicts.
I emphasize that these objects are not only pressure-supported, 
but they are also nearly
isothermal; i.e., there is not a large variation in the line-of-sight
velocity dispersion 
across these objects.  Stars are also pressure-supported systems but 
they would lie far outside the upper left boundary of this plot.  
However, stars are very far from isothermal.  

It has been noted above that, with MOND,
such self-gravitating near-isothermal systems would be expected to have 
internal accelerations comparable to or less than $a_o$.  
{\it But it is not at all 
evident how Newtonian theory can account for the fact these  
different classes
of astronomical objects, covering a large range in size and located in 
very different
environments, all appear to have comparable internal accelerations
near the cosmologically interesting value of $cH_o$.}

With MOND, systems that
lie below the line, i.e., with low internal accelerations, would be
expected to  
exhibit larger discrepancies.  This is particularly true of the
dwarf spheroidal systems.  Systems above the line (ellipticals) are
high surface brightness systems and if interpreted in terms of
Newtonian dynamics, would not exhibit much need for dark matter 
inside an effective
radius.  This seems to be the case.  I just add that the MOND
M-$\sigma$ relation (eq.\ 7) is very sensitive to variations from
strict homology which would be expected to lead to a large scatter in
the observed Faber-Jackson law.  However, MOND imposes boundary conditions on 
the inner Newtonian solution which restrict non-homologous objects to 
lie on a narrow fundamental plane similar to that implied by the 
traditional virial theorem (Sanders 2000).

Note that clusters of galaxies lie below the ${\sigma_l}^2/r = a_o$
line in Fig.\ 6;  thus, these objects would be expected to exhibit
significant discrepancies.
That this is the case has been known for 70 years (Zwicky 1933),
although the subsequent discovery of hot X-ray emitting gas 
goes some way in alleviating the original discrepancy.  For an isothermal
sphere of hot gas at temperature T, the Newtonian dynamical 
mass within radius
$r_o$, calculated from the equation of hydrostatic equilibrium, is 
$$M_n = {{r_o}\over G} {{kT}\over m} 
\Bigl({{d\,ln(\rho)}\over{d\,ln(r)}}\Bigl),\eqno(8)$$
where $m$ is the mean atomic mass and the logarithmic density gradient
is evaluated at $r_o$.  For the X-ray clusters plotted in Fig.\ 6 this 
turns out to be typically about a factor of 4 or 5 larger than
the observed mass in hot gas and in the stellar content of the galaxies.
This rather modest discrepancy viewed in
terms of dark matter has led to the so-called baryon catastrophe-- not
enough non-baryonic dark matter in the context of standard
CDM cosmology (White et al. 1993).  

With MOND, the dynamical mass (eq.\ 6) is
given by $$M_m = {(Ga_o)}^{-1} {\Bigl({{kT}\over m}\Bigr)^2}
\Bigl({{d\,ln(\rho)}\over{d\,ln(r)}}\Bigl)^2,\eqno(9)$$ and
the discrepancy, using the same value of $a_o$ determined from nearby galaxy 
rotation curves, is on average reduced to about a factor of 2 
larger than the observed mass.  There does indeed 
seem to be a remaining discrepancy.  This could be interpreted as a failure, 
or one could say that MOND predicts that
the baryonic mass budget of clusters is not yet complete and that there is
more mass to be detected (Sanders 1999). It would have certainly
been devastating for
MOND had the predicted mass turned out to be typically {\it less} 
than the observed mass in hot gas and stars.

\section{Cosmology and structure growth.}

Let me just summarize what I have said so far.  MOND not only allows the
form of rotation curves to be precisely predicted from the 
distribution of observable matter, but it also explains certain systematic
aspects of the photometry and kinematics of galaxies and clusters:  the
presence of a preferred surface density in spiral galaxies and ellipticals--
the so-called Freeman and Fish laws; the
fact that pressure-supported nearly isothermal systems ranging from
molecular clouds to clusters of galaxies are characterized
by a specific internal acceleration ($a_o$);
the existance of a TF relation with small scatter-- specifically a
correlation between the baryonic mass and the asymptotically flat 
rotation velocity of the form $v^4\propto M$;
the Faber-Jackson relation for ellipticals, and with more detailed 
modeling, the Fundamental Plane;  not only the magnitude of the discrepancy
in clusters of galaxies but also the fact that mass-velocity dispersion 
relation which applies to elliptical galaxies (eq.\ 6) extends to clusters
(the mass-temperature relation).  And it accomplishes all of this with a
single new parameter with units of acceleration-- a parameter determined from
galaxy rotation curves which is within an order of magnitude of the
cosmologically significant value of $cH_o$.  This is why several of us believe
that, on an epistemological level, MOND is more successful than dark matter.  

But, of course, MOND must fit into a larger picture.
One may naturally ask-- what are the larger-scale implications of modified 
dynamics-- specifically  what are implications for gravitational
lensing and does MOND imply a reasonable cosmology and cosmogony?
These are questions which require a more basic theory 
underlying MOND, and this is, at present, the essential weakness of
the idea.  

Frequently, the absence of a covariant theory is presented
as an argument against MOND.  But the criterion for judging a scientific
hypothesis surely must be its empirical success.  
The absence of a successful covariant
version is simply an aspect of its incompleteness.  People don't reject
general relativity because there is not yet a viable theory of quantum 
gravity.  At the same time,
it is fair to say that MOND will never be entirely credible to 
most astronomers and physicists until it 
makes some contact with more familiar physics.  
\begin{figure}
\begin{center}
\epsfig{file=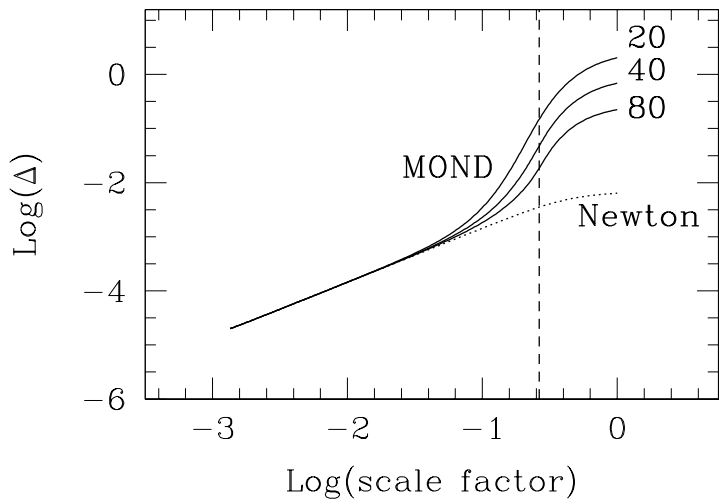,width=10cm}
\caption{The growth of fluctuations with an initial
amplitude of $10^{-5}$.  The solid lines show the growth of fluctuations on
various comoving scales in the context of the simple non-relativistic 
MOND theory (Sanders 2001) and dotted line is the usual Newtonian growth 
in the pure baryonic Universe.  The vertical dashed line indicates the scale
factor at which the cosmological term begins to dominate the expansion in
this model universe.}
\end{center}
\end{figure}

There have been
several attempts to construct a more general theory, most notably by
Bekenstein (1987), and while these are very nice ideas, none of 
these attempts is entirely satisfactory for various reasons
(Bekenstein \& Sanders 1994, Sanders 1997).  
A different approach is to consider MOND as modified
inertia (Milgrom 1994), perhaps resulting from the interaction of 
an accelerating particle with vacuum fields (Milgrom, 1999).  
Here the coincidence between $a_o$ and $cH_o$ plays a central role:  if
inertia results from influence of the vacuum on accelerated motion,
then, because a cosmological constant has a non-trivial effect upon
the vacuum, we might expect that it also has a non-trivial effect
upon inertia.  It is beyond my mission to describe these ideas in
detail, but
I would just like to comment upon the possible shape of a MOND cosmology.

First of all, I take it that the experimental foundations of the standard
Big Bang are so well-established, that any underlying theory of MOND
should not lead to a radically different cosmology, at least not in
the early Universe.  Then, to say that
MOND is an alternative to dark matter does not mean that 
every baryon in the Universe must be glowing with an M/L of one.  
In fact this is certainly not the case since $\Omega$ in visible matter is 
substantially less than $\Omega$ in baryons.  Moreover, 
there are clear indications that
at least some flavors of neutrinos have a non-vanishing mass (J. Bahcall,
this conference), so there is
a contribution of non-baryonic dark matter to the total mass
budget of the Universe-- at least at a level comparable to the mean 
density of baryons in visible stars.  But it would be contrary to the spirit 
of MOND if dark matter-- baryonic or non-baryonic-- were a dominant
constituent of the Universe or of bound gravitational systems such as galaxies
or clusters of galaxies.  So the question arises-- is cosmology with
$\Omega_m \approx \Omega_b \approx 0.02/h^2$ compatible with observations--
in particular, with the recent Boomerang and Maxima observations of the CMB
fluctuations?  

This is a question that has been considered by McGaugh (1999, 2000),
who applied the widely-used CMBFAST program (\cite{sz96}) 
in the case of a pure baryonic
Universe.  Before the Boomerang and Maxima results appeared (Hanany et al.
2000, Lang et al. 2001), McGaugh pointed
out that a pure baryonic universe, with the dominant constituent of the
Universe being in vacuum energy density, would imply that the second
peak in the angular power spectrum should be much reduced with respect
to the expectations of the concordance $\Lambda$CDM cosmology.  
The reason for
this suppression is basically Silk damping (Silk 1968)
in a low $\Omega_m$, pure baryonic universe--
the shorter wavelength fluctuations are exponentially suppressed by
photon diffusion.  When the Boomerang results appeared, much
of the excitement was generated by the unexpected low amplitude of the
second peak.  With $\Omega_{total} = 1.01$ and $\Omega_m=\Omega_b$ 
(no CDM or non-baryonic matter of any sort) McGaugh produced a rather
nice match to the Boomerang results.  A further prediction is
that the third peak should be even more reduced.  There are indications
from the recent more complete analyses of BOOMERANG data
(Netterfield et al. 2001) that this may not be the case, 
but the systematic uncertainties remain large.  
In addition, the SNIa results on the accelerated expansion of the 
Universe (Perlmutter et al. 1999) as well
as the statistics of gravitational lensing (Falco et al. 1998) seem to
exclude a pure baryonic and vacuum energy Universe, although it is unclear
that all systematic effects are well-understood.  It is also possible that
a MOND cosmology may differ from standard Friedmann cosmology in the low-z 
Universe (note that in some brane-world scenarios late-time cosmology 
diverges from Friedmann cosmology, e.g., Deffayet 2001).
\begin{figure}
\begin{center}
\epsfig{file=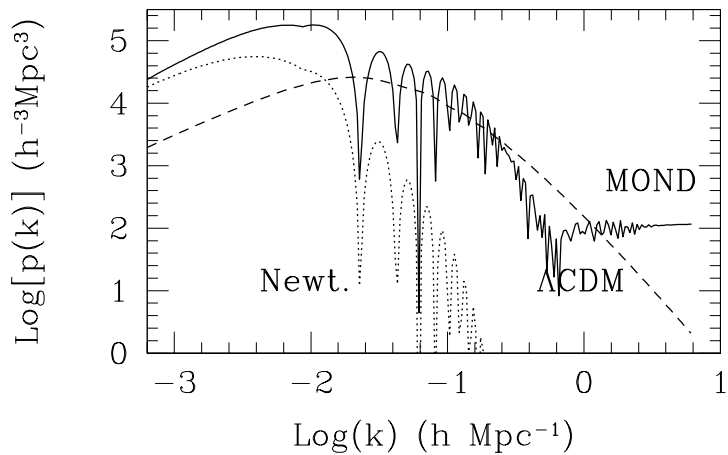,width=10cm}
\caption{The solid line shows the the power spectrum resulting from the
toy MOND theory applied to the growth of fluctuations with an initial 
Harrison-Zelodvitch, COBE-normalized power spectrum.  The dashed line
is the usual $\Lambda$CDM power spectrum and the dotted line shows the
power spectrum that would result from the Newtonian growth of fluctuations
in the pure-baryonic Universe}
\end{center}
\end{figure}

Of course, if we live in a Universe of only baryons, then how does 
structure form?  After all a primary motivation for non-baryonic
cosmic dark matter is the necessity of forming the observed structure
in the Universe by the present epoch via gravitational growth of very
small density fluctuations.  As we all know,
non-baryonic dark matter helps because it offers the possibility 
that fluctuations can begin growing before the epoch of hydrogen recombination.
The expectation is that MOND, by providing stronger  
effective gravity in the limit of low accelerations, might also help.

In the absence of a proper theory, this question can be considered by
making several {\it Ans\"atze} in the spirit of the existing bits of the
theory:

1.  The MOND acceleration parameter $a_o$ is constant with cosmic time.
This could be the case if $a_o$ is related to the cosmological term
($a_o \approx c\sqrt{\Lambda}$).
 
2.  MOND is applied in determining the peculiar accelerations-- those 
accelerations which develop around density perturbations-- and not
to the overall Hubble flow.  That is to say, the Hubble flow remains
intact.  One might imagine that MOND should be applied to the  
Hubble flow;
that is, as soon as the deceleration of the Hubble flow over a finite
size region falls below $a_o$, then the dynamics of that region begins
to deviate from the standard Friedmann solutions.  This would lead to
the eventual collapse of any finite size region regardless of its
initial density and expansion velocity (Felten 1984, Sanders 1998).  
With this sort of cosmology
the evolution of the early Universe would be as it is in the standard
Big Bang (the deceleration on relevant scales is much larger than $a_o$)
but the present Universe would look rather different than it actually does.

3.  Although the MOND does not affect the Hubble flow, the deceleration or
acceleration of the Hubble flow enters as a background field which influences
the development of peculiar accelerations in the MOND regime.

It is possible to construct a non-relativistic Lagrangian-based theory which
incorporates these three assumptions (Sanders 2001)-- this is similar 
to the 2-field version of the theory of Bekenstein and
Milgrom (1984) .  Following the same procedure
as in Newtonian cosmology, I find a growth equation for small density
fluctuations which is non-linear even in the regime where the density
fluctuations are small (this is because MOND is fundamentally
non-linear).  The growth of
fluctuations becomes dramatically rapid when the non-linear term 
dominates as is evident in Fig.\ 7 which is a plot of the fluctuation 
amplitude as a function of scale factor in a baryonic-vacuum energy
dominated Universe.  Fluctuations of smaller wavelength grow to larger
amplitude because they enter the MOND regime earlier.

The non-linear term becomes important  
when the background acceleration vanishes, i.e., when
the density in vacuum energy becomes comparable to the matter energy density.
Thus in a MOND Universe we might expect structure and massive galaxies
to form about when the cosmological constant begins to dominate
the expansion.
Starting with an initial Harrison-Zeldovitch power spectrum normalized
by COBE, the final power spectrum is shown in Fig.\ 8 where it is compared to
the $\Lambda$CDM power spectrum.  We see that it is quite similar 
apart from the baryonic oscillations.

So MOND offers the possibility of overcoming the slow growth of fluctuations
in a pure baryonic Universe.  It also offers an explanation of why we are
observing the Universe at an epoch when $\Lambda$ has only recently
emerged as the dominant term in the Friedmann equation.  If this scenario
remains as an aspect of a fully covariant theory, then the
cosmological argument is no longer a unique rationale for non-baryonic 
dark matter.

\begin{acknowledgements}
I am very grateful to Stacy McGaugh for useful discussions and for
preparing Fig.\ 5.  I thank Mario Livio for his kind invitation
to speak at this conference and for his efforts in organizing such an 
excellent and well-balanced scientific program. 
\end{acknowledgements}

\end{document}